\documentclass[12pt]{article}

\begin{document}

\title{General trends of the  late period of evolution in the
quasichemical model of nucleation}
\author{Victor Kurasov}

\maketitle

\begin{abstract}
The periods after the end of the "primary" nucleation are considered.
The approximate analytical description is given. The process is
split  into several periods which form the loop of evolution.
\end{abstract}

\section{Pleminary remarks}

The power of metastability in the system is described by the value
of supersaturation $\zeta$ defined as the ratio of surplus number
density
of molecules to the saturated number density of molecules.
After the end of nucleation one can show that the size droplet
spectrum is rather well localized in the space of sizes.
If we choose the free molecular regime of growth (in order to
escape from the  problems of the density profiles around droplets)
it is natural  to come to the natural size coordinate
$$
\rho = \nu^{1/3}
$$
where $\nu$ is the number of molecules inside the droplet. The
law of regular growth is rather simple
$$
\frac{d \rho}{dt} = \frac{\zeta}{\tau} (1- \frac{\rho_c}{\rho})
$$
with the characteristic time $\tau$
as a parameter and the critical size $\rho_c$
which is determined as
$$
\rho_c = 2a / 3 \ln(\zeta+1)
$$
where $a$ is the renormalized surface tension.
For the small supersaturations
$$
\rho_c  \approx 2a / 3 \zeta
$$
On the base of the critical size the value of supersaturation can
be expressed as
$$
\zeta = 2a / 3 \rho_c
$$

The asymptotic rate of growth for the supercritical droplets is
very simple
$$
(\frac{d \rho}{dt})_{as}
= \frac{\zeta}{\tau}
$$

From the
theory of nucleation it is known that the distribution function is
well localized in the scale of sizes. Namely, the half-width
$\Delta_1 x$ in $\rho$-axis satisfies the following estimate
$$
f_* (\Delta_1 x)^4 \simeq \Gamma^{-1} \Phi_*
$$
Here
$f_*$ is the amplitude of spectrum,
$\Phi_*$ is the "quantity" of surplus vapor
calculated in terms of supersaturation,
parameter $\Gamma$ can be
estimated as
$$
\Gamma \simeq \nu_c
$$
and the value $\nu_c$ is the number of molecules in the critical
embryo.

The number of droplets can be estimated as
$$
N = f_* \Delta_1 x
$$

At the final of the surplus vapor exhaustion one can write
$$
N z_{fin}^3 = \Phi_*
$$
where
$z_{fin}$ is an imaginary final coordinate in $\rho$-scale (actually
there is no real final coordinate it is some quasistationary
approximation).

We consider the situation of decay here but the situation of
dynamic conditions can be easily considered by quasistationary
approximations and by the first step of corresponding iteration
procedures.

Then we come to
\begin{equation} \label{1}
\frac{z_{fin}}{\Delta_1 x} = \Gamma^{1/3}
\sim \nu_c^{1/3}
\end{equation}

Since $\nu_c \gg 1$ we see the certain hierarchy.

\section{Regular relaxation of the size spectrum}

When the asymptotic law  of the
regular growth is used
the evolution of the system after the end of nucleation \cite{TMF}
can be described by means of the first order differential equation
$$
\tau \frac{dz}{dt} = \Phi_*
- \sum_i \mu_i^{\infty} z^{3-i} C_3^i
$$
with "coefficients" $\mu_i^{\infty}$ as the full momentums of the
distribution function\footnote{With appropriate renormalization.}
$$
\mu_i^{\infty} = \int_{-\infty}^{\infty}  x^i f(x) dx
$$
Here $f(x)$ is the distribution density determined by the
nucleation theory \cite{TMF}. The variable $x$ is the shift at
$\rho$-scale of the size of droplet from the "marked" size $z$ of
a droplet born at some moment.

If we choose as $z$ the size of the droplet formed at the initial
moment of time then
$$
\mu_i^{\infty} = \int_{0}^{\infty}  x^i f(x) dx
$$

The initial condition for the differential equation is the
following: At
$ t=t_* + \Delta_1 t $ the
position of the size spectrum is $ z=\Delta_1 x
$.
Here $t_*$ is the "characteristic" moment, $\Delta_1 t$ is the
duration of the "back" side spectrum formation.
Approximately $\Delta_1 t = \Delta_1 x \tau / \Phi_*$. So, $t_* +
\Delta_1 t$ is the moment of the "end" of the nucleation period.
Also here $\Delta_1 x$ is the length of the back side of the size
spectrum. All these characteristics can be given by the nucleation
theory.

This differential equation can be easily integrated
$$
\tau \int \frac{dz}{\Phi_* -
\sum_i \mu_i^{\infty} z^{3-i} C_3^i } = t
$$

Condition of the end of the regular relaxation is
\begin{equation}\label{t}
( 2 \div 3)
\rho_c =
z- \Delta_1 x
\end{equation}
or
$$
( 2 \div 3)
\rho_c =
z
$$
Since
$$
\rho_c = \frac{2a}{3 \ln(\zeta+1) } \approx
\frac{2a}{3 \zeta }
$$
and
$$
\zeta =\Phi_*
- \sum_i \mu_i^{\infty} z^{3-i} C_3^i
$$
it is an ordinary algebraic equation which can be effectively
solved on the base of the strong inequality
$$\zeta \ll \Phi_*$$

At the moment when
the equation
(\ref{t}) is satisfied one has to take into account the
diffusion correction to the law of droplets growth.

The final values can be made more accurate. We can use the balance
equation
$$
\Phi = \frac{2a}{3\rho_0} + N \rho_0
$$
to get $\rho_0$.

\section{Diffusion erosion  of spectrum}

The process of diffusion leads to destruction of the previous form of spectrum.
Since the spectrum is rather narrow and (\ref{1}) takes place one
can use initial condition as $\delta$-function type. But it is
not absolutely  necessary.

There are two approaches which can be used here. At first one can
assume that the spectrum is narrow according to (\ref{1}) and use
the Green function in $\nu$-axis
$$
G = \frac{1}{\sqrt{4 \pi D}} \exp(\frac{ - (\nu - \nu_0)^2}{4Dt})
$$
Here
$\nu_0 = \rho_0^3$,
$$
D = (W^+ + W^-)/2
$$
$W^+$ is absorbtion coefficient and $W^-$ is desorbtion
coefficient. These coefficients are taken at $\nu = \nu_0$.

Another possibility is to take $\nu_0$ as the mean coordinate
corresponding to $z^3$ in the previous solution of the relaxation
problem.

The effect of dependence of $D$ on $\nu$ leads to the following
equation
$$
\frac{\partial f}{\partial t} = - D_0 \nu^{2/3}
\frac{\partial^2 f}{\partial \nu^2}
$$
where
$$
D_0 = D / \nu_0^{2/3}
$$
does not essentially depend on $\nu$.
The last equation can be approximately written as
$$
\frac{\partial f}{\partial t} = - D_0
\frac{\partial^2 f}{\partial S^2}
$$
for
$$
S = \nu^{2/3}
$$
Then the solution is the Green function  and it is known
\begin{equation} \label{sol}
G = \frac{N}{\sqrt{4 \pi D_0}} \exp(\frac{-(S - S_0)^2}{4D_0t})
\end{equation}
Here
$$
S_0 = \nu_0^{2/3}
$$
Until the half-width $4Dt$ is many times less than $\nu_0$ there is
no difference what variable ($\nu$ or $S$) is used.

One can remark that the initial form of the size spectrum will
disappear rather soon. Really, the characteristic width of
spectrum is
$$
\Delta_2 S =
(z+\Delta_1 x)^2 - z^2 \simeq 2 \Delta_1 x z
$$
and
$$
\frac{\Delta_2 S}{S} = 2 \frac{\Delta_1 x }{z}
$$

After the time $t_2$ the diffusion erosion of the size spectrum
will have the scale of the initial width of spectrum and the
initial form of the size spectrum completely disappears. This time
allows the estimate
$$
t_2 = (\Delta_2 S)^2 /(4D)
$$
Since $\Delta_1  x / z_{fin} \ll 1$ one can use both $x$ or $S$
representation of solution.

One has to note that in the process of relaxation $\zeta$ varies
and $D = D(\zeta)$ varies. The law $\zeta(t)$ is already known from
solution of the regular evolution equation.
 Then one can suggest the following approximate account
 of variation of $D$: instead of $Dt$ in denominator
one can approximately use
$\int D(\zeta(t)) dt$.

One has also to  refine the regular evolution equation, which allows to prolong it. Really,
having written the law of growth
$$
\frac{dz}{dt} =
\frac{\zeta}{\tau} ( 1 - \frac{\frac{2a}{3 \zeta}}{z})
$$
with
$$
\zeta = \Phi - \sum_i \mu_i^{\infty} z^{3-i}  C_3^i
$$
This differential equation will be more accurate and complex.
The problem is that $\mu_i^{\infty}$ begin to depend on $t$ and $z$.

But here fortunately appears the monodisperse approximation.
Since $\Delta_1 x \ll z_{fin}$
then
$$
z^3 \mu_0^{\infty} \approx N z^3\gg
z^2 \mu_1^{\infty} \approx N  \Delta_1 x z^2 \gg
z \mu_2^{\infty} \approx N  \Delta_1^2 x z \gg
 \mu_3^{\infty} \approx N  \Delta_1^3 x
$$
So, one can use the monodisperse approximation.
The monodisperse approximation is rather accurate. In this
approximation
$$
\zeta = \Phi - N z^3
$$
Then
$$
\frac{dz}{dt} =
\frac{\Phi - N z^3}{\tau} (1-\frac{2a}{3\Phi z - 3 Nz^4} )
$$
The main momentum is $\mu_0^{\infty}$. The main advantage is that
 this  momentum doesn't change in the real regular law of growth
 instead of the asymptotic one.
 The last equation can be easily integrated
$$
\int
\frac{dz}{(\Phi - N z^3) (1-\frac{2a}{3\Phi z - 3 Nz^4} )} =
\frac{t}{\tau}
$$

Since $0<dz/dt< \zeta/ \tau$ for $z>z_c$ one can see the "narrowing"
of the spectrum.

Now we have to extract the limits of a "pure diffusion".
One has to take into account that in reality the pure diffusion
equation takes place only in the "extended near-critical zone". While
the
boundary of the "near critical zone"  is determined by condition
$$
F_c-F \simeq  1
$$
the "extended near-critical zone" (ENZ) is determined by two conditions
$$
\frac{dF}{d\nu} \ll 1
$$
$$
\nu < (2 \div 3) \nu_c \equiv \beta^3 \nu_c
$$
The second condition is necessary for the case $\zeta \ll 1$ to
exclude the infinite zone of big $\nu$.
If we require that the diffusion length $\Delta S \sim \sqrt{4Dt}$
has to be compared with the regular length $\Delta S \sim \zeta^2
t^2 / \tau^2 $, then we come to $\rho^3 \sim 4/\zeta$, which is
less than $\rho_c = 2a/3\zeta$.

Certainly, since now
$dz/dt$ depends on $z$ the values of $\mu_i$ will change. So, here
there are only two possibilities: 1). to consider $\mu_i$
slowly varying; 2). to solve everything in the monodisperse
approximation.

One can see that both "near-critical zone" and "extended
near-critical zone" are  not symmetrical in respect to $\nu_c$.
This requires to use more sophisticated methods instead of the
leading term of the steepens descent to calculate the Zeldovitch'
factor. But this can be considered as technical details.

The lower boundary where the first condition fails even for small
$\zeta$ depends on the concrete value of surface tension. We
denote the boundary as
$$
\nu_l = \nu_c \alpha \ \ \ \ \  \rho_l = \rho_c \alpha^{1/3}
$$

The value of $\alpha$ can be estimated when we require that the
derivative of the free energy equals by the absolute value to the
derivative of the free energy for big supercritical embryos. This
leads to $\alpha = 1/8$.

The spread of the gaussian occurs up to the moment
$$
t_3 = (1-\alpha^{2/3})^2 S_c^2  /(4D_0)
$$
where
$S_c$ corresponds to the final coordinate.

When the droplet attains $\nu_l$ then rather soon (in comparison with the time
of attaining $\alpha \nu_c$) it will be
dissolved. The
probability of growth back is small.

When the essential part of spectrum is in the ENZ one can use the
presented solution.
One can observe and prove the relaxation to
the quasi-equilibrium.  In
quasiequilibrium the behavior of $\rho_c$ is regulated by  the
balance condition
$$
\int f(\rho,t) 3 \rho^2 \frac{d\rho}{dt} d\rho = 0
$$
($f(\rho,t)$ is the known distribution)
or by
$$
\int f(\rho,t) 3 \rho^2 \frac{\zeta}{\tau}
(1-\frac{\rho_c}{\rho}) d\rho = 0
$$
It is important that the vapor consumption by droplets with a known spectrum of sizes
 regulates
the behavior of $\rho_c$ contrary to the method of Lifshitz-Slezov
\cite{LS},
where the behavior of $\rho_c$ regulated the form of spectrum.

Actually nothing happened when $f(\rho,t)$ is narrow. Let it be
like $\delta$-function $\delta(\rho_a)$. Then $\rho_a$ relaxes to
$\rho_c$ and the process stops until the diffusion makes spectrum
wider.

\section{Further evolution of spectrum in a near critical region }

On the base of solution (\ref{sol}) one can see the transformation  of
spectrum. The spectrum is like a Gaussian and according to the
balance condition $\rho_c$ is near the head of the Gaussian.
Then it is clear that at first the lower boundary of ENR will be
attained by the essential part of spectrum. The upper boundary can
not be attained first.

We see that the  essential part of spectrum is localized in the
region where $|\nu-\nu_c| \ll 1$. Then
at first there is no difference
whether to use the diffusion equation in $\nu$-scale or in
$S$-scale.

The time when the essential part of spectrum attains $\rho_l$ can
be easily calculated
$$
4 D t = (\nu_c - \nu_l)^2
$$
Here we calculate time from the moment of the end of the previous
period.

After this moment we have to take into  account that practically all droplets
at $\rho_l$ are going to dissolve. Then
$$
f(\rho)|_{\rho=\rho_l} = 0
$$

Solution of diffusion equation (it has to be solved now in
$S$-scale) with such a boundary condition is
$$
f(\rho) = 0
$$
for
$$
\rho < \rho_l
$$
and
\begin{equation} \label{sol1}
f(S) = G_+ - G_-
\end{equation}
for
$$
S^{1/2} > \rho_l
$$
Here
$$
G_+ =
\frac{N}{\sqrt{4\pi D t}} \exp( -
\frac{(S-S_0)^2}{4 D t}
)
$$
$$
G_+ =
\frac{N}{\sqrt{4\pi D t}} \exp( -
\frac{(S-S_1)^2}{4 D t}
)
$$
and $S_0$ and $S_1$ are defined by
conditions that $S_0 = z_{fin}^2$ and $S_0 - S_l = S_l - S_1$
(where $S_l = \nu_l^{2/3}$).

A special question appears what is here $D$. But at first we
shall analyze the properties of solution. One can see that
contrary to the unbounded solution with a half-width
$$
\Delta S \sim t^{1/2}
$$
and the maximum (and mean) value is staying
$$
S_{max} = const
$$
the new solution
has  the moving maximum  value
$$
S_{max} \sim t^{1/2}
$$
the moving mean value
$$
S_{mean} \sim t^{1/2}
$$
and the growing half-width
$$
\Delta S \leq t^{1/2}
$$
which  ensures the relative localization of spectrum.

So, we have the following estimate
$$
\Delta S \leq S_{mean}
$$

The value of supersaturation has to be determined from
\begin{equation} \label{difbal}
\int_0^\infty d\rho 3 \rho^2 f(\rho) \frac{d\rho}{dt} = 0
\end{equation}
Then we get $\rho_c$ and the supersaturation.

Then in this solution we have to take into account that
$D\approx D(\rho_c(\zeta))$.
Again we can approximately
substitute $Dt$ by $\int D(\zeta(t)) dt $.

Actually, this is the end of solution. Everything presented later
refers to rather small tails which can be seen only in
enormously big systems.

\section{Formation of a tail}

The previous  equation to get the supersaturation is not a unique
way to do it.  Rigorously speaking  the
supersaturation has to be derived
 from the substance balance equation
\begin{equation} \label{intbal}
\frac{2a}{3 \rho_c} = \Phi_*  - \int_0^\infty \rho^3 f(\rho,t) d\rho
\end{equation}
Distribution has at first to be taken here from (\ref{sol1}).

One can see that the part of the size spectrum with big $\rho$ is
important in the substance balance due to the factor $\rho^3$.
But the law of growth of the droplets of big sizes is regular, it
is not the diffusion walking on the flat potential surface as it
was supposed in derivation of solution (\ref{sol1}).

We suppose that at
$$
\rho_r = \beta \rho_c
$$
the law of growth will be changed to the regular asymptotic law of
growth
$$
\frac{d\rho}{dt} = \frac{\zeta}{\tau}
$$
Here $\beta$ is a parameter, the normal value is
$$
\beta = 1.5
$$

The rate of appearance of droplets at the beginning of the region
of regular growth can be calculated as following. At first  we can
calculate $f(\rho)$ by
$$
f(\rho) = f(S) \frac{dS}{d\rho} = f(S) 2 S^{1/2}
$$
on the base of solution (\ref{sol1}).
Then
$$
J= f(\beta \rho_c) \frac{\zeta}{\tau}
$$
or
$$
J= f(\beta \rho_c) \frac{\zeta}{\tau}(1-\beta)
$$
for the intensity of formation of the tail.
Then we get the tail formed according to
$$
f_{tail} (\rho,t) = f(\beta \rho_c(t'))\frac{\zeta(t')}{\zeta(t)}
$$
where
$t'$ is defined as
$$
\rho= \beta \rho_c + \int_{t'}^t \frac{\zeta(t'')}{\tau} dt''
$$

Let $z$ be a coordinate $\rho$ of some droplet\footnote{More convenient is to choose
$z$ as the coordinate of the front side of the tail.}. Then $x$ is
$z-\rho$ for the given droplet. It remains constant during the
regular growth. Then
$$
f_{tail} (\rho,t) = f_{tail} (x)
$$
 and it is
a known function.

One can see the following approximation.
Certainly, the position of $S_c$ coincides in initial moments of
time with $S_0$ and do not vary essentially. It a quasi-integral of
evolution. Then denoting
$$
\zeta_0 = \frac{2a}{3\rho_{0}}
$$
one can put $\zeta_0$ instead of $\zeta(t'')$ in the previous
formula. The movement of the boundary $S_r = (\beta \rho_c)^2$ can be neglected and one
can act in a quasi-stationary approximation. As the
result one gets the form of the tail.

More accurate is to take $S_c$ from the solution (\ref{sol1})
ignoring the formation of the tail. Then
$$
\zeta_0 (t'') = \frac{2a}{3S_c^{1/2} (t'')}
$$

Now we can present the
distribution as
$$
f = f_{\pm} + f_{tail}
$$
and in the quasiequilibrium we have
\begin{equation}
\int_0^\infty d\rho 3 \rho^2 (  f_{\pm} + f_{tail} )
\frac{d\rho}{dt} = 0
\end{equation}
This gives $\rho_c$ in quasiequilibrium and a new value of
supersaturation. This closes the iteration loop. Actually only one
loop is necessary.

\section{Dissolution of the head of spectrum}

At first the tail is not important in the substance balance. But
the quantity of substance $G_{tail}$ in the tail grows faster than
something proportional to
$t^3$. So the droplets in the tail become the essential consumers
of vapor.

Now the distribution density is the following
$$
f(\rho) = 0
$$
for
$$
\rho < \rho_l
$$
and
\begin{equation}
f(S) = G_+ - G_-
\end{equation}
for
$$
\rho_r > S^{3/2} > \rho_l
$$
$$
f= f_{tail}
$$
for
$$
\rho>\rho_r
$$
Then one can get the  supersaturation from
the balance equation
$$
\zeta = \Phi_* - \int_{S_l}^{S_r} f(S) S^{3/2} dS +
\int_{\rho_r}^\infty f_{tail}(\rho) \rho^3 d\rho
$$

At first the vapor consumption of the droplets in the tail i.e.
the evolution of
$$
G_{tail} (t) =
\int_{S_r}^{\infty}  f_{tail} (\rho, t) d\rho
$$
determines the value of supersaturation
$$
\frac{d}{dt} \zeta = - \frac{d}{dt} G_{tail}
$$
The head of the spectrum begins to dissolve since the droplets in
the head of spectrum  become mainly subcritical. This diminishes
the quantity
$$
G_{head} = \int_{S_l}^{S_r} f(\rho,t) dt
$$
and partially stabilizes the supersaturation.

The result will be the dissolution of the head of the size
spectrum. The final of this period will be when
$$
N_{tail} \sim N_{total} - N_{tail}
$$

One can propose the approximate (not necessary)  model: at first
the intensity of vapor consumption by the tail can be calculated
according to
$$
\frac{d}{dt} G_{tail} =
3 \int_{\rho_r}^\infty f_{tail} (\rho) \rho^2 \frac{d \rho}{dt} d\rho
$$

Then the number $N_{head}$ in the head satisfies the  following
equation
$$
[\frac{d}{dt} N_{head} ] z_c^3 + \frac{d}{dt} G_{tail} = 0
$$
The growth of $G_{tail}$ occurs in the avalanche manner. So, the
dissolution of the head occurs in the avalanche manner also.
So, this solution can prolonged up to
$$
N_{tail} \ll N - N_{tail}
$$

\section{Dissolution of the essential tail of spectrum}

One can spread the solution
to the beginning of the dissolution of the tail of spectrum.

But when the
complete dissolution of the head is finished one can propose the
more effective procedure.

From
$$
\frac{2a}{3 (z-x_c)} = \Phi - \sum_i  C_3^i  z^{(3-i)} \mu_i (z)
$$
where the changing momentums are
$$
\mu_i = \int_{-\infty}^{\tilde{x}} x^i f_{tail}(x) dx
$$
(here $\tilde{x}$ corresponds to $\rho = \beta \rho_c$).
are known functions of $x_c$ we take $x_c$ as a function of $z$.
The region $\rho < \beta \rho_c$ is negligible in definition of
$\mu_i$ when $z>>\rho_c$.
Now the law of growth
$$
\tau \frac{dz}{dt} =
\frac{2a}{3 (z-x_c)}
$$
becomes the closed differential equation of the first order.
Since
the r.h.s. of the last equation is known function of $z$ the last
equation can be easily integrated
$$
t = \tau  \int \frac{dz}{\frac{2a}{3 (z-x_c(z))}}
$$
It can be rewritten as
$$
\tau \frac{dz}{dt} = \Phi - \sum_i  C_3^i  z^{(3-i)} \mu_i (z)
$$
Since
$
\mu_i(\tilde{x}(z))
$
can be well approximated as polynomials on $z$, the integration
can not cause difficulties.

Here one can see the important property which allows to solve this
equation very simply: the vapor is consumed by the droplets with
regular growth.

One can show the simplicity of solution by the following example:
suppose that the height of tail $f_*$ is constant. Then
$$
G_{tail} =  f_* (z^4 - z_c^4\beta^4)
$$
and
$$
\tau \frac{dz}{dt} =
\Phi - f_* (z^4 - z_c^4 (z) \beta^4)
$$
One can integrate this equation. Rewrite this equation for
evolution of $z_c(t)$
$$
\frac{2a}{3z_c} = \Phi + f_* z_c^4 \beta^4 - f_* z^4
$$
Then $z$ can be expressed through $z_c$
$$
z= f_*^{-1/4}
(\Phi + f_* z_c^4 \beta^4 - \frac{2a}{3z_c} )^{1/4}
$$
Having differentiated the last relation and substituted it into
the initial balance equation
$$
\Phi - f_*(z^4 - f_* z_c^4 \beta^4) =
\tau \frac{1}{3z^3} \frac{dz^4}{dt}
$$
we get
$$
 \frac{2a}{3\tau z_c} =
\frac{1}{3} f_*^{-1/4} (\Phi +  f_* z_c^4 \beta^4  -
\frac{2a}{3z_c})^{-3/4} (f_* 4 z_c^3 \beta^4 + \frac{2a}{3z_c^2})
\frac{dz_c}{dt}
$$
which can be integrated
$$
t  = \tau
\int
\frac{3z_c}{2a}
f_*^{-1/4}
(f_* 4 z_c^3 \beta^4 + \frac{2a}{3z_c^2}) \frac{1}{3}
(\Phi + f_* z_c^4 \beta^4 - \frac{2a}{3z_c})^{-3/4} dz_c
$$

The end of this period takes place when the number of droplets in
ENR, namely $N_{ENR}$ becomes comparable with the rest droplets
$$
N_{ENR} \sim N_{total} - N_{ENR}
$$

\section{Closure of the loop of evolution }

Later one
has to take into account the diffusion of the rest of the tail.
Again we
have the spectrum of some form in ENR.
Since the form of the  tail is rather flat we take the flat
spectrum with a gaussian front side.
At first we see the dissolution of the flat part of the size
spectrum. It is easy to describe
because we see the one dimensional heat
conductivity problem with  a boundary
condition of a following type: at $S=S_l$ the distribution $P$ is
zero; at big $S$ the distribution $P$ vanishes, then at $S = z^2$
one can see the free diffusion. The initial condition is
$$
P (x, t=0) =
\Theta(S-S_l) A \Theta(S_k - S)
$$
where
$A$ is the amplitude of tail (constant or slowly varying function)
 and $S_k$ is the position of the
front side of tail (in shifted coordinates), $\Theta(x)$ is the Heavisaid's function.

To keep condition $P=0$ at $S=S_l$ we can prolong the previous
initial condition oddly.

The solution of diffusion problem can be found anywhere.

Having moved in $S$-scale the origin to the former $S_l$ we see
the solution
$$
P  =
\int F(x,0) exp(-\frac{(S-x)^2}{4Dt} ) dx
$$
Here $F(x,0)$ is the distribution
given by
$$
F(x,t=0) =
\Theta(S)\Theta(S_k-S_l -S)A -
\Theta(-S)\Theta(S+S_k-S_l) A
$$
It is quite clear that when $A$ is  constant
or can be approximated by polynomials then
the solution can be written via
a linear combination of error
functions.

It is also clear that when $$t  \gg  S_k^2/(4D)$$ then the solution begin
to resemble the gaussian.

When the flat
region is dissolved the size spectrum resembles the gaussian and
we come to the already studied problem.
 So the procedure of all
sections starting from the diffusion erosion of the size spectrum in the
ENR has to be repeated. This closes the loop and these loops will be inevitably
continued in future. One has to stress that the number of droplets
radically decreases in every loop and to see these loops one has
to observe an extremely giant system.
To give description on the base of continuous approximation
even in the beginning of the tail of the size spectrum there have
to be many droplets.
Earlier or later  diffusion and stochastic growth of droplets lead
to the change of the regime of evolution presented here.
Also the change of regime of vapor consumption, the heat release
effects, coagulation etc. destroy the presented solution.

\end{document}